\documentclass[doublecol]{epl2}
\usepackage{graphicx}

\title{Generalized partial measurements}

\author{G.~S.~Paraoanu}
\institute{Low Temperature Laboratory, Aalto University, P. O. Box 15100, FI-00076 AALTO, Finland.}

\abstract{
We introduce a type of measurements that generalize the so-called "partial measurements" performed in recent years with
phase qubits. While in the case of partial measurements it has been demonstrated that
one could undo the effect of the measurement only for non-switching events, we show here that generalized partial measurements can be reversed probabilistically for both switching and non-switching events. We calculate the associated Fisher information
and discuss the estimation sensitivity for quantum tomography. Two ways of implementing this type of
measurements with superconducting qubits are proposed.
}

\begin{document}
\maketitle

\section{Introduction}

We consider a type of measurements which are generalizations of the "partial measurements"  demonstrated experimentally in recent years in the field of superconducting qubits \cite{partialmartinis}. An interesting feature of partial measurements is that they can be reversed  \cite{katz} if the SQUID used for measurement did not switch in the running-wave state. In case the SQUID has switched, the qubit is destroyed, and a voltage is registered by  the measuring device \cite{me},  precisely like in quantum optics an event would correspond to a photon being absorbed in a detector. In contradistinction,
 our generalized partial measurements have the property that they can be probabilistically reversed for both measurement results (switch or non-switch). We further show that such measurements can be performed along any direction in space and we calculate the corresponding Fisher information metric. Finally, we give two explicit physical constructions of these measurements, one using two qubits and the other using a single qubit in a two-well potential.

\section{Generalized partial measurements: definition}
We consider a qubit with states $|0\rangle$ and $|1\rangle$, prepared in an unknown pure state $|\psi \rangle = \cos (\theta /2) |0\rangle + \exp (i\varphi ) \sin (\theta /2) |1\rangle $. Partial measurements \cite{partialmartinis} are characterized by a single parameter $p$, which is the probability of switching (or tunneling)  from the state $|1\rangle$; tunneling from the state $|0\rangle$ is forbidden, and as a result one of the Kraus operators defining these measurements is in fact a projection. The immediate generalization of partial measurements is to allow also $|0\rangle$ to switch (tunnel) with probability $q$. This results in POVM  measurements \cite{povm} described by two measurement operators, $M_{m}$ and $M_{\bar{m}}$,
\begin{eqnarray}
M_{m} &=& \sqrt{1-q}|0\rangle\langle 0| + \sqrt{1-p}|1\rangle\langle 1| , \label{embarg} \\
M_{\bar{m}} &=& \sqrt{q}|0\rangle\langle 0| + \sqrt{p}|1\rangle\langle 1| , \label{mg}
\end{eqnarray}
and corresponding to two effects $E_{m}=M_{m}^{\dagger}M_{m}$ and $E_{\bar{m}}=M_{\bar{m}}^{\dagger}M_{\bar{m}}$, which are positive operators realizing a semispectral resolution of the identity, $E_{m} + E_{\bar{m}} = 1$. Here $m$ and $\bar{m}$ ("not m") are measurement results corresponding respectively to the absence or existence of a switching (tunneling) event. As mentioned before, the parameters $0\leq p \leq 1 $ and $0\leq q \leq 1$ have the meaning of probabilities of obtaining the result $\bar{m}$  depending on whether the qubit is in the state $|1\rangle$, or  $|0\rangle$ respectively: indeed, the probabilities of obtaining the results $m$ and $\bar{m}$ can be immediately calculated,
\begin{eqnarray}
P_{m} &=& \langle \psi |E_{m}|\psi\rangle =1- q\cos^2 (\theta/2)  - p\sin^2 (\theta/2)  ,\label{unog}\\
P_{\bar{m}}&=& \langle \psi |E_{\bar{m}}|\psi\rangle = q \cos^2 (\theta/2)  + p \sin^2 (\theta/2),
\label{duog}
\end{eqnarray}
and the wavefunctions after the measurement are $|\psi_{m}\rangle$ and $|\psi_{\bar{m}}\rangle$,
\begin{eqnarray}
|\psi_{m}\rangle &=& \frac{1}{\sqrt{P_{m}}}M_{m}|\psi\rangle  \label{afterg}\\
&=&
\frac{1}{\sqrt{P_m}}\left[\cos \frac{\theta}{2} \sqrt{1-q}|0\rangle  +
\sin \frac{\theta}{2}
\sqrt{1-p}e^{i\varphi} |1\rangle\right],\nonumber \\
|\psi_{\bar{m}}\rangle &=& \frac{1}{\sqrt{P_{\bar{m}}}}M_{\bar{m}}|\psi\rangle \label{aftershaveg}\\
&=&\frac{1}{\sqrt{P_{\bar{m}}}}\left[\cos \frac{\theta}{2} \sqrt{q}|0\rangle + \sqrt{p}e^{i\varphi} \sin \frac{\theta}{2}|1\rangle \right].\nonumber
\end{eqnarray}

\section{Reversal of generalized partial measurements}
We now turn to the problem of time-reversing the measurements: we claim that, if $p,q \notin \{0,1\}$, both measurements defined by Eqs. (\ref{embarg}, \ref{mg}) can be probabilistically reversed, thus bringing back the system to the exact initial state. Indeed, if $p,q \notin \{0,1\}$, $M_{m}$ and $M_{\bar{m}}$ admit an inverse,
\begin{eqnarray}
M_{m}^{-1}&=& \frac{1}{\sqrt{(1-p)(1-q)}}XM_{m}X , \\
M_{\bar{m}}^{-1} &=& \frac{1}{\sqrt{pq}}XM_{\bar{m}}X ,
\end{eqnarray}
where here and in the following we will use the standard quantum-information notations \cite{nielsen} for the Pauli matrices,
\begin{equation}
X=\left(\begin{array}{cc} 0&1 \\1 &0\end{array}\right),
 Y=\left(\begin{array}{cc} 0&-i \\i &0\end{array}\right),
Z=\left(\begin{array}{cc} 1&0 \\0 &-1\end{array}\right). \nonumber
\end{equation}
Suppose now we were measuring  an arbitrary state $|\psi\rangle$ and the result $m$ has occurred: then, we apply $X$ and measure again, and if the result $m$ occurs again, then we apply X again and recover $|\psi\rangle$; if, instead, we get the result $\bar{m}$  then we fail. The other possibility is that, when we first measure $|\psi\rangle$,
the result $\bar{m}$ occurrs: we apply $X$, measure again, and again either $\bar{m}$ occurs, in which case we apply $X$ and recover $|\psi\rangle$; or, $m$ occurs, in which case we fail. The probability of success in each case can be found by using  the usual rules of multiplication for conditional probabilities at each step or directly by  $\langle \psi |M_{m}^{-1}XM_{m}X |\psi\rangle = (1-p)(1-q)$ for the path $m\rightarrow m$
 and $\langle \psi |M_{\bar{m}}^{-1}XM_{\bar{m}}X |\psi\rangle = pq$ for the path $\bar{m}\rightarrow\bar{m} $, giving a total success rate of $1-p-q+2pq$, independent of the state.

One might now wonder if it is possible to reverse the result of the measurement deterministically: that is, is it possible to reverse also the "fail" results
$M_{\bar{m}}XM_{m}|\psi\rangle  = M_{m}XM_{\bar{m}}|\psi\rangle = \sin(\theta /2)e^{i\varphi }|\sqrt{q(1-p)}|0\rangle + \cos(\theta /2) \sqrt{p(1-q)}|1\rangle$
in the scheme above? Up to a global phase factor, any $2\times 2$ unitary can be written as
$R_{z}(\alpha )R_{y}(\beta )R_{z} (\gamma )$, where $R_{y}$, $R_{z}$ are rotations
around the  axes $y,z$, $R_{z}(\alpha ) = \exp (-i \alpha Z/2)$, $R_{y}(\beta ) = \exp (-i \beta Y/2)$. The reversibility condition is $R_{z}(\alpha )R_{y}(\beta )R_{z} (\gamma )M_{\bar{m}}XM_{m}|\psi\rangle = |\psi\rangle$; after some algebra, we find that this  implies $\alpha =\beta =\gamma =0$ (that is, an $X$ gate again), and also $p=q$. This means that the only situation when we can have deterministic reversal (reversal of all paths) is the trivial case in which the measurement operators Eqs. (\ref{embarg},\ref{mg}) become identity. In this situation we also notice that Eqs. (\ref{unog},\ref{duog}) yield $P_{m}=P_{\bar{m}} =1/2$ with no state dependence, meaning that no information about $\theta$ can be obtained! Of course,
$M_{m}$ and $M_{\bar{m}}$ being the identity effectively signifies that no measurement has been done. This proves that it is not possible to reverse deterministically a generalized partial measurement.


\section{Quantum tomography using generalized partial measurements: Fisher information}
 Next, we show that generalized partial measurements can be performed along any direction, once they are experimentally  available along the $z$ direction, as in Eqs. (\ref{embarg}, \ref{mg}). Indeed, given an arbitrary direction $\vec{n}=(n_{x}, n_{y}, n_{z})$, parametrized in spherical coordinates ($n_{x}=\sin\chi \cos\psi$,
 $n_{x}=\sin\chi \sin\psi$, and $n_{z}=\cos\chi$), we can always find a rotation that
 brings $|0\rangle$ to $|+\rangle_{\vec{n}}$ and
 $|1\rangle$ to $|-\rangle_{\vec{n}}$. Here $|\pm\rangle_{\vec{n}}$ are eigenvectors of the spin operator along the direction $\vec{n}$, $(\vec{n}\vec{\sigma})|\pm\rangle_{\vec{n}} = \pm |\pm\rangle_{\vec{n}}$. Then,
by performing a rotation before and after applying the measurement operators Eqs. (\ref{embarg},\ref{mg}), we get the new measurement operators along $\vec{n}$,
\begin{eqnarray}
M_{m}^{(\vec{n})} &=& \sqrt{1-q}|+\rangle_{\vec{n}\vec{n}}\langle +| + \sqrt{1-p}|-\rangle_{\vec{n}\vec{n}}\langle -| , \label{embarggen} \\
M_{\bar{m}}^{(\vec{n})} &=& \sqrt{q}|+\rangle_{\vec{n}\vec{n}}\langle +| + \sqrt{p}|-\rangle_{\vec{n}\vec{n}}\langle -| , \label{mggen}
\end{eqnarray}
Similarly to Eqs. (\ref{unog}, \ref{duog}) we get  the effects $E^{(\vec{n})}_{m}=M^{(\vec{n})\dagger}_{m}M^{(\vec{n})}_{m}$ and $E^{(\vec{n})}_{\bar{m}}=M^{(\vec{n})\dagger}_{\bar{m}}M^{(\vec{n})}_{\bar{m}}$, as well as the probabilities
\begin{eqnarray}
P^{(\vec{n})}_{m} &=& \langle \psi |E^{(\vec{n})}_{m}|\psi\rangle = \langle \psi |M^{(\vec{n})\dagger}_{m}M^{(\vec{n})}_{m}|\psi\rangle
, \label{oss}\\
P^{(\vec{n})}_{\bar{m}} &=& \langle \psi |E^{(\vec{n})}_{\bar{m}}|\psi\rangle =
\langle \psi |M^{(\vec{n})\dagger}_{\bar{m}}M^{(\vec{n})}_{\bar{m}}|\psi\rangle
. \label{osss}
\end{eqnarray}

We can now calculate the Fisher information matrix, defined in general for a conditional probability distribution $P({\zeta}|\Theta)$ (where $\zeta$ stands for the measurement outcomes and $\Theta$ a vectorial parameter)  as the  average ${\cal F}_{\Theta_{i},\Theta_{j}} = \langle [\partial_{\Theta_{i}} \ln P(\zeta|\Theta)]\cdot [\partial_{\Theta_{j}} \ln P(\zeta |\Theta )]\rangle$ over the values $\zeta$ (here $\partial_{\Theta_{i}} =\partial /\partial \Theta_{i}$). The Fisher information is a measure of the sensitivity of our measurement scheme to the determination of $\theta$ and $\varphi$, thus it is directly relevant for quantum tomography.
In our case, we have
$\Theta = (\theta, \varphi )$, $\zeta\in\{m,\bar{m}\}$, giving
a binary distribution $P^{(\vec{n})}_{m}+P^{(\vec{n})}_{\bar{m}}=1$. This yields
\begin{eqnarray}
{\cal F}^{(\vec{n})}_{\theta, \varphi} &=& {\cal F}^{(\vec{n})}_{\varphi ,\theta} = \frac{1}{P_{m}^{(\vec{n})}P_{\bar{m}}^{(\vec{n})}} [\partial_{\theta} P_{m}^{(\vec{n})}]
[\partial_{\varphi} P_{m}^{(\vec{n})}], \label{f1}\\
{\cal F}^{(\vec{n})}_{\theta, \theta}  &=& \frac{1}{P_{m}^{(\vec{n})}P_{\bar{m}}^{(\vec{n})}}\left[\partial_{\theta}  P_{m}^{(\vec{n})} \right]^2 ,\label{f2}\\
{\cal F}^{(\vec{n})}_{\varphi, \varphi}  &=& \frac{1}{P_{m}^{(\vec{n})}P_{\bar{m}}^{(\vec{n})}}\left[\partial_{\varphi} P_{m}^{(\vec{n})} \right]^2.\label{f3}
\end{eqnarray}

\begin{figure}[t]
\begin{center}
  \includegraphics[width=7.5cm]{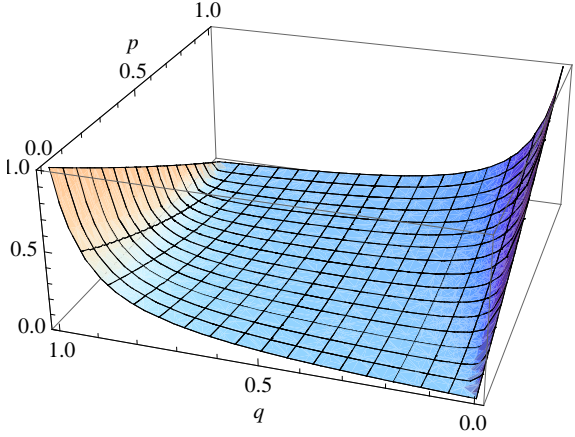}
\end{center}
\caption{Plot of the function ${\cal F}_{\theta\theta}$ for $\chi =0$ and $\theta =\pi/6$ as a function of $p$ and $q$.}
         \label{fig1}
\end{figure}

For the derivatives, we get
\begin{eqnarray}
\partial_{\theta} P_{m}^{(\vec{n})}&=& \frac{q-p}{2}[\sin\theta \cos\chi -\cos\theta\sin\chi\cos (\varphi -\psi )], \nonumber\\
\partial_{\varphi} P_{m}^{(\vec{n})}&=& \frac{q-p}{2}\sin\theta\sin\chi\sin (\varphi -\psi ) \nonumber.
\end{eqnarray}
These expressions, together with Eqs. (\ref{oss}, \ref{osss}), provide analytical expressions for calculating the Fisher matrix elements Eqs. (\ref{f1}-\ref{f3}) of a generalized measurement along any direction and for any state. As an example, in Fig. (\ref{fig1}) we take $\chi =0$, making all the elements of the Fisher matrix zero except for ${\cal F}^{(\vec{n})}_{\theta, \theta}$, which is plotted as a function of $p$ and $q$ for $\theta = \pi/6$.

Generalized partial measurements are subject to the usual uncertainty principle: the precision in determining one variable is limited by the Cram\'{e}r-Rao bound ({\it e.g.} $(\Delta \theta)^2 \geq 1/{\cal F}_{\theta\theta}$, if $\varphi$ is known, or in general $\langle (\Theta -\tilde{\Theta})(\Theta -\tilde{\Theta})^{t}\rangle \geq {\cal F}^{-1}$, where $\tilde{\Theta}$ is the true value of $\Theta$).

In addition, another type of complementarity can be noticed. First, we see that no matter in which direction we perform the measurement, the four elements of the Fisher information matrix depend on $(p-q)^2$ (see {\it e.g.} Fig. \ref{fig1}). Thus, the analysis based on the Fisher information confirms that no knowledge about the parameters $\theta, \varphi$ can be acquired near $p=q$.
 Therefore, to be able to do quantum tomography using generalized partial measurements it is preferable to have $p$ and $q$ as different from each other as possible (that is, one close to zero and the other one close to 1).  But, on the other hand,  we see that attempting to determine precisely the parameters characterizing the  state
increases the risk of not being able to reverse the measurement in either one of the two paths described above.
Indeed, the likelihood of successful reversal on a path  $m\rightarrow m$ is $(1-q)(1-p)$, while on the path $\bar{m}\rightarrow \bar{m}$ is $pq$, and maximization of one leads to the other being zero.
Now, since we have no knowledge about the state of the system, it is not possible to decide which path should be optimized; we can attempt to  optimize them together
by looking at the conditional entropy for successful measurements followed by reversals, $-(1-p)(1-q) \ln (1-p)(1-q) - pq \ln pq$. At $p=q=1/2$ this quantity reaches its maximum value of $\ln 2$, and  at this point the Fisher information is zero.

\section{Physical implementation}

Physically, a POVM measurement can be regarded as a von Neumann measurement on another system (called "ancilla") which has previously interacted with the qubit, with the result described by a unitary $U$. The existence of this transformation $U$ is guaranteed by Naimark's dilation theorem but its realization in relation with experiment is usually not straightforward. Mathematically, given the basis $|0,m\rangle$, $|0,\bar{m}\rangle$,
$|1,m\rangle$, $|1\bar{m}\rangle$, we can construct a unitary
\begin{equation}
U = \frac{I + Z}{2}\otimes U_{q} +  \frac{I - Z}{2}\otimes U_{p}.
\end{equation}
For example, for $U_{q} = \sqrt{1-q} I + i \sqrt{q} Y$ and $U_{p} = \sqrt{1-p} I + i \sqrt{p} Y$ we can check that the measurement operators $M_{m}$ and $M_{\bar{m}}$ are obtained by applying $U$ and performing sharp (von Neumann) measurements with projections $|m\rangle\langle m|$ and $|\bar{m}\rangle\langle \bar{m}|$. Note also that even more general forms of the measurement operators can be obtained by using, instead of the $U_{q}$'s above, the most general form of a unitary on $2 \times 2$ matrices.
\begin{figure}[t]
\begin{center}
  \includegraphics[width=7cm]{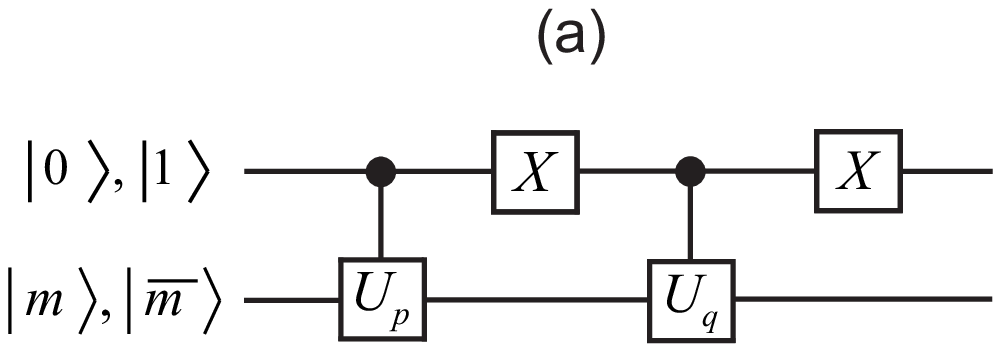}\hfill\includegraphics[width=6cm]{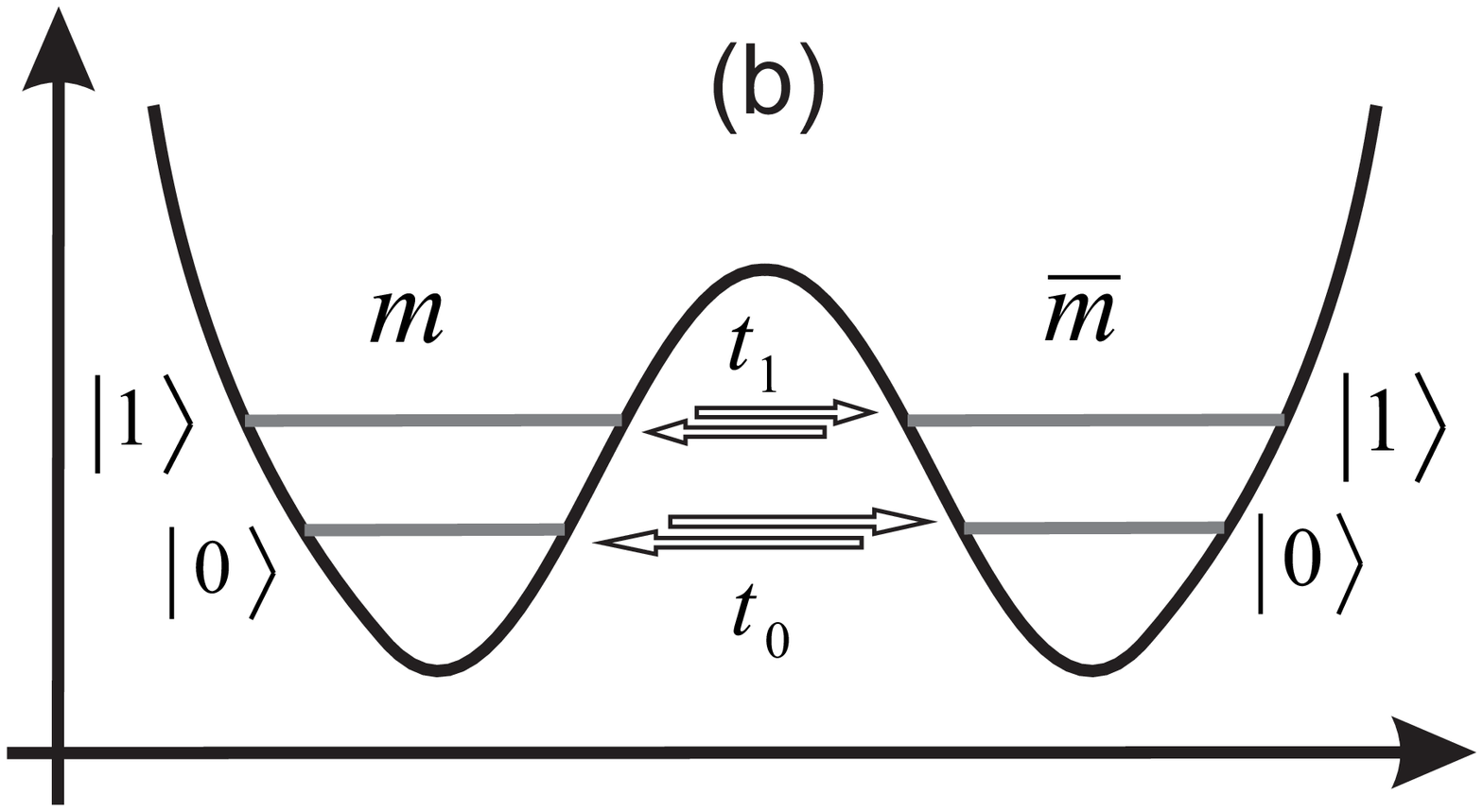}
\end{center}
\caption{(a) Implementation of the unitary $U$ as a sequence of gates. (b) Schematic of a double-well implementation of the unitary $U$ in a double-well potential. The vertical axis is an energy scale and the horizontal axis is the coordinate (a macroscopic superconducting phase difference, in the case of superconducting qubits).}
         \label{fig2}
\end{figure}

To experimentally realize $U$, a straightforward implementation is to simply consider two qubits and assume that one has available a universal set of gates. In Fig. \ref{fig2}(a) we show the expansion of $U$ in single-qubit $X$ gates and two-qubit conditional gates. Such gates are currently available in many experimental setups, including
superconducting qubits (see {\it e.g.} \cite{thesis} for the implementation of the conditional gates from Fig. \ref{fig2}(a) for flux qubits).

Although possible in this way, it is still useful to look for an implementation {\it via} a single operation, which would save significantly on the time required to perform the measurement.
The operation time can be an important restriction, since many qubits have a limited coherence time. Such a physical implementation can be reached if
we regard the ancilla as another degree of freedom of the same qubit. In this way, the number of qubits required is only one, which is significant if one weighs in the technical difficulties of controlling two qubits and their coupling.
Consider for example a symmetric double-well potential with adjustable barrier height (see Fig. \ref{fig2}(b)). Initially, the barrier is high and the qubit is prepared in the left well in a superposition of the states $|0\rangle$ and $|1\rangle$ with energy difference $\hbar \nu$, then it is lowered for some time $\tau$, allowing the qubit to tunnel, and then it is raised again.
The tunnel matrix elements $t_0$ and $t_1$ depend ({\it via} the WKB formula) on the energy (therefore on the state) of the qubit. We assume that the energy scales $\nu \gg t_{0}, t_{1}$ are such that the lowering and raising of the barrier is adiabatic with respect to the qubit energy levels but instantaneous with respect to the tunneling rates. We refer to the left and right wells as $m$ and $\bar{m}$ respectively.
The Hamiltonian of the system during  the time $\tau$ is
\begin{equation}
H= -\frac{\hbar \nu}{2}Z\otimes I +
t_{0}\frac{I + Z}{2}\otimes X + t_{1}\frac{I - Z}{2}\otimes X .
\end{equation}
We then solve  the evolution problem in the frame rotating at the qubit frequency, {\it i.e.} we transform the wavefunction by $|\psi\rangle \rightarrow \exp (- i\nu Z\otimes I /2) \tau  |\psi\rangle$, obtain the new Hamiltonian, and evolve during $\tau$ to finally get the evolution operator
\begin{eqnarray}
U &=& \frac{I + Z}{2}\otimes [\cos\frac{t_{0}\tau}{2} I- i \sin \frac{t_{0}\tau}{2}X] + \nonumber\\
&&+\frac{I - Z}{2}\otimes [\cos\frac{t_{1}\tau}{2} I - i \sin \frac{t_{1}\tau}{2}X] .\nonumber
\end{eqnarray}
This obviously produces the required $U$ with the identification $q= \sin^2(t_{0}\tau /2\hbar )$ and $p= \sin^2(t_{1}\tau /2\hbar )$.

Then, one should have available the projections $|m\rangle\langle m|$ and $|\bar{m}\rangle\langle \bar{m}|$. In the field of superconducting qubits, only the first operator was available in standard switching-current measurements: the projection $|m\rangle\langle m|$ was realized if the measuring junction or SQUID did not switch in the running-wave state during the time $\tau$. If it did switch, the qubit was destroyed (similar to photon absorbtion with optical qubits) due to quasiparticle generation at the point when the voltage reaches the superconducting gap. Recently, it was shown that QND measurements that could implement $|\bar{m}\rangle\langle \bar{m}|$ can be realized if the voltage is prevented to reach the gap by fast-feedback control electronics \cite{mooij}. Thus, with such devices it could be possible to implement the measurement operators described above.

\section{Conclusion}

We introduced a class of POVM measurements that generalizes the partial measurements demonstrated experimentally with superconducting qubits. We show that for these measurements it is possible to reverse (probabilistically) both of the measurement results. We study the associated Fisher information and we propose a physical implementation of these measurements.

\acknowledgements
Financial support from the Academy of Finland is acknowledged (grant 00857, and projects 129896, 118122, and 135135).

\end{document}